\documentclass{llncs}
\usepackage{amsfonts}
\usepackage{epsfig}
\usepackage{color}
\usepackage{amssymb}
\usepackage{latexsym}
\usepackage{enumerate}
\usepackage{amsmath}
\usepackage{graphicx}
\def\alert{\emph}
\def\adimH{\mathsf{K}} 
\def\bmat{\begin{pmatrix}}
\def\emat{\end{pmatrix}}

\def\C{\mathbb{C}} 

\def\GAP{\textit{GAP}}
\def\Hspace{\mathcal{H}} 
\def\id{\mathbf{1}} 
\def\idmat{\mathrm{I}} 
\def\ig{\gamma} 
\def\iG{\Gamma} 
\def\iGN{{\cabs{\iG}}} 
\def\iGX{\iG^{\X}} 
\def\ls{\sigma} 
\def\lS{\Sigma} 
\def\lSN{{\cabs{\lS}}} 
\def\lSX{\lS^{\X}} 
\def\natmod{\mathsf{H}} 
\def\N{\mathbb{N}}
\def\NF{\mathcal{F}}

\def\period{\mathcal{P}} 
\def\Prob{\mathrm{\mathbf{P}}} 
\def\Q{\mathbb{Q}} 
\def\regrep{\mathrm{P}} 

\def\repp{\mathrm{P}} 
\def\repq{\mathrm{U}} 
\def\repqco{\mathrm{V}} 
\def\runisymb{\mathsf{r}} 
\def\sg{\mathsf{f}} 
\def\sG{\mathsf{F}} 
\def\sGN{{\cabs{\sG}}} 
\def\tin{0} 
\def\tfin{T} 
\def\Time{\mathcal{T}} 
\def\transmatr{\mathrm{T}} 
\def\wg{\mathsf{g}} 
\def\wG{\mathsf{G}} 
\def\wGN{\mathsf{M}} 
\def\ws{\omega} 
\def\wS{\Omega} 
\def\wSN{\mathsf{N}} 
\def\x{\mathsf{x}} 
\def\X{\mathsf{X}} 
\def\XN{{\cabs{{\X}}}} 
\def\Z{\mathbb{Z}}
\newcommand{\AltG}[1]{\mathsf{A}_{#1}} 
\newcommand{\Aut}[1]{\mathrm{Aut}\vect{#1}} 
\newcommand{\barket}[1]{\left|#1\right\rangle} 
\newcommand{\cabs}[1]{\left|#1\right|} 
\newcommand{\class}[1]{K_{#1}} 

\newcommand{\inner}[2]{\left\langle#1\mid#2\right\rangle} 
\newcommand{\invar}[3]{\mathrm{#1}\vect{#2,#3}} 
\newcommand{\invarL}[2]{\mathrm{L}_{#1}\vect{#2}} 
\newcommand{\invarQ}[3]{\mathrm{Q}_{#1}\vect{#2,#3}} 
\newcommand{\IrrRep}[1]{\mathbf{#1}} 
\newcommand{\Math}[1]{$#1$} 
\newcommand{\Mathh}[1]{$$#1$$} 
\newcommand{\Mone}[1]{\bmat#1\emat} 
\newcommand{\Mthree}[9]{\bmat#1&#2&#3\\
 #4&#5&#6\\
 #7&#8&#9\emat} 
\newcommand{\Mtwo}[4]{\bmat#1&#2\\#3&#4\emat} 
\newcommand{\Nei}[1]{\mathrm{N}\vect{#1}} 
\def\Oppbare{{^\mathrm{c}}}
\newcommand{\Opp}[1]{{#1}\Oppbare}
\newcommand{\ordset}[1]{\left[#1\right]} 
\newcommand{\Perm}[1]{\mathrm{Sym}\left(#1\right)} 
\newcommand{\PermRep}[1]{\mathbf{\overline{#1}}} 
\newcommand{\ProbBorn}[2]{\Prob\!\vect{#1,#2}} 
\newcommand{\runi}[1]{\runisymb_{#1}} 
\newcommand{\set}[1]{\left\{#1\right\}} 
\newcommand{\SymG}[1]{\mathsf{S}_{#1}} 
\newcommand{\vect}[1]{\left(#1\right)} 
\newcommand{\Vthree}[3]{\bmat#1\\#2\\#3\emat} 
\newcommand{\Vtwo}[2]{\bmat#1\\#2\emat} 
\begin{document}
\title{Mathematical Modeling of Finite Quantum Systems}
\titlerunning{Finite Quantum Systems}
\author{Vladimir V. Kornyak}
\institute{Laboratory of Information Technologies \\
           Joint Institute for Nuclear Research \\
           141980 Dubna, Russia \\
           \email{kornyak@jinr.ru}}
\authorrunning{Vladimir V. Kornyak}
\maketitle
\begin{abstract}
We consider the problem of quantum behavior in the finite background. 
Introduction of continuum or other infinities into physics leads only to 
technical complications without any need for them in description of empirical observations.
The finite approach makes the problem constructive and more tractable.
We argue that quantum behavior is a natural consequence of symmetries 
of dynamical systems. It is a result of fundamental impossibility to 
trace identity of indistinguishable objects in their evolution --- 
only information about invariant combinations of such objects is available.
We demonstrate that any quantum dynamics can be embedded into a simple permutation
dynamics. Quantum phenomena, such as interferences, arise in invariant subspaces of  
permutation representations of the symmetry group of a system. Observable quantities  
can be expressed in terms of the permutation invariants.
\end{abstract}
\section{Introduction}
Unitary operators in Hilbert spaces --- in fact, unitary representations of some 
symmetry groups --- lie in the core of mathematical descriptions of quantum phenomena.
We can assume finiteness of these groups without any risk to destroy
the physical content of the problem, since metaphysical choice between
``\emph{finite}'' and ``\emph{infinite}'' can not lead to any empirically observable
consequences. Moreover, there are strong experimental evidences that 
{finite groups of}
relatively \emph{small orders} underlie some fundamental physical processes
\cite{RevPartPhys,Ishimori,Ludlgen,HPS02,HS03,Altarelli,Smirnov,BlumHagedorn}.
These evidences come mainly from the flavor physics, especially in the lepton sector.
The origin of these small groups is unclear in the context of currently 
accepted theories like, e.g., the Standard Model.
\par
Let us recall some standard facts from the group theory \cite{HallEn}:
\begin{itemize}
	\item 
Any linear representation of a finite group \Math{\wG=\set{\wg_1,\ldots,\wg_{\wGN}}} 
is \emph{unitary}.
	\item  
Any irreducible representation is contained in the \emph{regular representation}
--- matrix version of permutation action of the group  \Math{\wG} on 
its elements.
	\item  
Any set \Math{\wS=\set{\ws_1,\ldots,\ws_{\wSN}}} on which
\Math{\wG} acts transitively by permutations is equivalent to a set of cosets of some subgroup 
in \Math{\wG}.
\end{itemize}
Combination of these facts leads to the conclusion that any 
quantum problem with evolution operators 
\Math{\repq\!\vect{\wg_i}} 
belonging to a representation \Math{\repq} of \Math{\wG} in \Math{\adimH}-dimensional 
Hilbert space \Math{\Hspace_\adimH} can be reduced to permutations of 
\Math{\wSN\geq{}\adimH} things.
If \Math{\wSN>\adimH} then the permutation representation \Math{\repp} in
 \Math{\wSN}-dimensional 
Hilbert space \Math{\Hspace_\wSN} has the structure
\Math{\repp\cong\IrrRep{1}\oplus\repq\oplus\repqco}.
Additional ``\emph{hidden parameters}'' appearing due to increase of
Hilbert space dimension in no way can effect on the data relating to the space 
\Math{\Hspace_\adimH} since both this space and its complement 
are \emph{invariant subspaces}
in \Math{\Hspace_\wSN}. 
Quantum phenomena, such as, e.g., \emph{destructive interference}, 
can be observed in proper
invariant subspaces of the representation \Math{\repp}. 
More detailed examination --- starting with the fact that all eigenvalues 
of group representations are \emph{roots of unity} 
--- leads to the conclusion that 
in  ``finite'' quantum formalism the field of complex numbers should be 
replaced by some extension \Math{\NF} of rational numbers with abelian Galois group. 
This extension depends on the structure of underlying group \Math{\wG}. 
The field  \Math{\NF} is a subfield of some cyclotomic field 
(the \emph{Kronecker-Weber theorem}). Hermitian operators describing observables 
in quantum formalism are elements of representation of \emph{group algebra} 
over \Math{\NF}. 
All ingredients of quantum theory --- like, e.g., the Heisenberg \emph{uncertainty principle}
 --- are obtained in the standard way. Observable data in the space 
\Math{\Hspace_\adimH} can be expressed in terms of \emph{permutation invariants} 
relating to the space \Math{\Hspace_\wSN}.
\par 
The permutation representation \Math{\repp} makes sense in the \Math{\wSN}-dimensional
module \Math{\natmod_\wSN} over the semi-ring  \Math{\N=\set{0,1,2,\ldots}} 
of natural numbers. Introduction of \Math{\natmod_\wSN} is equivalent to prescribing
natural weights (``\emph{multiplicities of occurrences}'' or 
``\emph{population numbers}'') to the elements of underlying
set \Math{\wS}. The Hilbert space \Math{\Hspace_\wSN} is 
derived from  \Math{\natmod_\wSN}
by extending the semi-ring \Math{\N} to the field  \Math{\NF}. 
If quantum amplitudes in
\Math{\Hspace_\adimH} are obtained as projections of ``natural'' vectors from  
\Math{\natmod_\wSN\subset\Hspace_\wSN}, then the \emph{Born probabilities} in 
\Math{\Hspace_\adimH} appear to be \emph{rational numbers}. 
This is compatible with the \emph{frequency interpretation} of probability for finite sets:
the probability is the ratio
of singled out combinations to the total number of combinations under consideration.
The interpretational issues like ``\emph{wavefunction collapse}'',
``\emph{many-worlds}'', ``\emph{many-minds}'' etc. lose any meaning in the 
finite background.
The quantum phenomena arise as manifestation of fundamental impossibility to 
trace identity of indistinguishable objects in their evolution --- only information about
invariant combinations of such objects is available.
\section{Basic Structures}
We consider evolution of dynamical systems in the \emph{discrete time} \Math{\Time}.
We will assume that \Math{\Time=\Z} or \Math{\Time=\ordset{\tin,1,\ldots,\tfin}} 
for some \Math{\tfin\in\N}.
\par
\emph{Classical states} of dynamical system form a finite set 
\Math{\wS=\set{\ws_1,\ldots,\ws_{\wSN}}} of some entities.
\emph{Classical evolution} or \emph{trajectory} of the dynamical system is a sequence
 \Math{\ldots,\ws_{t-1},\ws_{t},\ws_{t+1},\ldots\in\wS^\Time}. 
 We assume the existence of a 
 \emph{symmetry group} \Math{\wG=\set{\wg_1,\ldots,\wg_{\wGN}}\leq\Perm{\wS}}
acting on the entities \Math{\wS}.
\par  
We will argue that \emph{quantum evolution} can be defined as
a sequence of permutations \Math{\ldots{}p_{t-1},p_{t},p_{t+1}\ldots},
where \Math{p_{t}=\wS{a_t}} is the permutation of the set \Math{\wS} by
a group element \Math{a_{t}\in\wG}.
A natural condition here is that at any moment \Math{t} only information
on invariant combinations of elements from \Math{\wS} is available, whereas the 
permutation \Math{p_t} itself is unobservable. More precisely, the information
about arrangement of elements from \Math{\wS} corresponding to \Math{p_t} does 
not make sense without a ``reference frame'' (or ``observer'').
\par
In physics the set \Math{\wS} usually has a special structure of a set of functions
\begin{equation}
\wS=\lSX
\label{stateswithspace}
\end{equation}
on a \emph{space} \Math{\X=\set{\x_1,\ldots, \x_\XN}} 
with values in a set  \Math{\lS=\set{\ls_1,\ldots,\ls_\lSN}} of \emph{local states}.
\par 
We assume that both the space \Math{\X} and the local states \Math{\lS}
possess  nontrivial groups of \emph{space}
\Math{\sG=\set{\sg_1,\ldots,\sg_\sGN}\leq\Perm{\X}}
and \emph{internal} \Math{\iG=\set{\ig_1,\ldots, \ig_\iGN}\leq\Perm{\lS}} 
\emph{symmetries}, respectively. In principle, the most general symmetry group
\Math{\wG} for the set of states having form \eqref{stateswithspace} can be
computed with the help of modern algorithms%
\footnote{The algorithm designed by B. McKay \cite{McKay}
is considered as the most efficient to date. Its main target are graphs, 
but its universal scheme can be adjusted without excessive efforts 
for computing symmetries of other combinatorial structures.}.
Of course, this might be combinatorially difficult task. However, we can easily 
construct some group \Math{\wG} directly from the groups \Math{\sG} and \Math{\iG}. 
The construction --- generalizing what is used in physical theories --- is the following
equivalence class of \emph{split extensions}
\begin{equation}
\id\rightarrow\iGX\rightarrow\wG\rightarrow\sG\rightarrow\id,
\label{extensionEn}
\end{equation}
where \Math{\iGX} is the group of \Math{\iG}-valued functions on the space \Math{\X}.
Any equivalence class is determined by an \emph{antihomomorphism} 
\Math{\mu: \sG\rightarrow\sG}. The term 
`antihomomorphism' means that \Math{\mu(ab)=\mu(b)\mu(a)} for \Math{a,b\in\sG}.
For example, choosing the \emph{trivial} \Math{\mu(a)=\id} and the \emph{natural} 
\Math{\mu(a)=a^{-1}} 
antihomomorphisms we obtain, respectively:
\begin{itemize}
	\item  the \emph{direct product}
\Math{\wG=\iGX\times\sG} (this is the standard choice in physical theories) 
	\item  the \emph{wreath product}
\Math{\wG=\iG\wr_\X\sG\equiv\iGX\rtimes\sG} (\Math{\rtimes} 
stands for the \emph{semidirect} product).
\end{itemize}
In \cite{KornyakNS10} we gave explicit formulas for group operations in \Math{\wG}
defined by \eqref{extensionEn} in terms of operations in \Math{\sG} and \Math{\iG}.
\par
An important property of dynamical systems with space%
is the presence of 
\emph{non-trivial} gauge connections. 
The gauge structures lead to observable 
physical consequences: the curvatures of non-trivial connections
describe forces in physical theories. Another important topic involving
the space structure is the spin/statistics relation. 
On the other hand there are many problems, for example, quantum computing, where
any underlying space is inessential.
\section{Finite Quantum Models}
As is well known, all approaches to quantization are equivalent to the traditional 
matrix formulation of quantum mechanics where the evolution of a system
from an initial to a final state is described by an \emph{evolution matrix} \Math{U}:
\Math{\barket{\psi_{\tin}}\rightarrow\barket{\psi_{\tfin}}=U\barket{\psi_{\tin}}}.
The evolution matrix of a quantum dynamical system can be represented as the product
of matrices corresponding to elementary time steps:
\Math{U=U_{\tfin\leftarrow\tfin-1}\cdots{}U_{t\leftarrow{}t-1}\cdots{}U_{1\leftarrow0}.}
We will follow the evolution matrix approach throughout this paper.
\par
The main ingredients of the standard quantum mechanics are the following:
\begin{enumerate}
	\item
Quantum description deals with \emph{unitary operators} \Math{U} acting
in a \emph{Hilbert space} \Math{\Hspace} over the field of complex 
numbers \Math{\C}. The elements \Math{\barket{\psi}\in\Hspace} of the space 
are called ``\emph{states}'', ``\emph{state vectors}'', ``\emph{wave functions}'', 
``\emph{amplitudes}'' etc.
The operators  \Math{U} belong to the general unitary 
group \Math{\Aut{\Hspace}} acting in \Math{\Hspace}.
	\item 
Quantum mechanical \emph{particles} are associated with
 \emph{unitary representations} in  \Math{\Hspace} of some symmetry groups.
The representations are called ``\emph{singlets}'', ``\emph{doublets}'',
``\emph{triplets}'' etc.
in accordance with their \emph{dimensions}.
The multidimensional representations describe  \emph{spin}
	\item
Quantum mechanical \emph{evolution} is an unitary transformation 
of the \emph{initial} state vector \Math{\barket{\psi_{in}}} into the \emph{final}
 \Math{\barket{\psi_{out}}=U\barket{\psi_{in}}}.
In the continuous time, an elementary step of evolution is described by
 the \emph{Schr\"{o}dinger equation}
\Mathh{\displaystyle{}i\frac{\mathrm{d}}{\mathrm{d}t}\barket{\psi}=H\barket{\psi},}
where \Math{H} is a \alert{Hermitian} operator called \alert{energy operator} or 
 \alert{Hamiltonian}. 
	\item
Quantum mechanical \emph{experiment} (\emph{observation}, ``\emph{measurement}'')
is a comparison of  the state  \Math{\barket{\psi}} of 
a \emph{system} with the state \Math{\barket{\phi}} of an \emph{apparatus}.
	\item
In accordance with the \emph{Born rule}, the \emph{probability} 
\Math{\ProbBorn{\phi}{\psi}} to register a particle 
described by  \Math{\barket{\psi}} by apparatus tuned to  
\Math{\barket{\phi}} is equal to 
\Math{\displaystyle\frac{\cabs{\inner{\phi}{\psi}}^2}
 {\inner{\phi}{\phi}\inner{\psi}{\psi}}}.
	\item
Quantum \emph{observables} are described by \emph{Hermitian operators} acting 
in the Hilbert space \Math{\Hspace}.
\end{enumerate}
Our aim is to reproduce all this in the constructive finite background.
Our strategy --- in accordance with the Occam principle --- will be to avoid introduction 
of entities unless we really need them.
Keeping these lines we come to the following:
\begin{enumerate}
\renewcommand{\labelenumi}{\theenumi$'$.}
	\item
The Hilbert space  \Math{\Hspace} over the field \Math{\C} should be replaced by a
\Math{\adimH}-dimensional Hilbert space \Math{\Hspace_\adimH} over an 
\emph{abelian number field} \Math{\NF} --- an \emph{extension} of the rationals
\Math{\Q} with an \emph{abelian Galois group} \cite{ShafarevichEn}. 
The unitary operators \Math{U} belong now to an unitary representation  \Math{\repq}
of a \emph{finite group} \Math{\wG=\set{\wg_1,\ldots,\wg_{\wGN}}} in the 
space \Math{\Hspace_\adimH}.
The field \Math{\NF} is determined by the structure of the group \Math{\wG}
and its representation \Math{\repq}.
	\item
The notion of \emph{quantum particle}	remains the same as in the standard
quantum mechanics. 
	\item 
It is clear	that now we have only finite number of possible evolutions:
\Mathh{U_j\in\set{\repq\vect{\wg_1},
\ldots,\repq\vect{\wg_j},\ldots,\repq\vect{\wg_\wGN}}.}
Obviously we do not need any analog of the  \emph{Schr\"{o}dinger equation} at all.
Though formally one can always introduce Hamiltonians by the formula
\Math{H_j=i\ln{}U_j\equiv\sum\limits_{k=0}^{p-1}\lambda_k{}U_j^k}, where \Math{p} is
period of \Math{U_j} (i.e., minimal \Math{p>0} such that \Math{U_j^p=\idmat}),
 \Math{\lambda_k}'s are some coefficients%
\footnote{These coefficients contain the \emph{non-algebraic} element 
\Math{\pi} which is an \emph{infinite} sum of elements from \Math{\NF}.
In other words, the \Math{\lambda_k}\!'s are elements of a \emph{transcendental extension}
of  \Math{\NF} --- the logarithmic function is essentially a
construction from the continuous mathematics dealing with infinities.}.
	\item
The notion of \emph{observation}	remains without changes.
	\item
The formula for Born's probability remains the same
\begin{equation}
\ProbBorn{\phi}{\psi} = \frac{\textstyle{\cabs{\inner{\phi}{\psi}}^2}}
{\textstyle{\inner{\phi}{\phi}\inner{\psi}{\psi}}}.
\label{BornEn}
\end{equation}
But some conceptual refinement is needed. In the finite background the only
reasonable interpretation of probability is the \emph{frequency interpretation}:
the probability is the ratio of the number of singled out combinations to the total 
number of combinations under consideration. So we expect that if all things are arranged
correctly, then formula \eqref{BornEn} must give \emph{rational numbers}.
We will use this as a guiding principle.
	\item
\emph{Hermitian operators} 
describing \alert{observables}
in quantum formalism can be expressed in terms of the \emph{group algebra} 
representation:
\begin{equation*}
	A=\sum\limits_{k=1}^{\wGN}\alpha_k{}\repq\vect{\wg_k}.
\end{equation*}
Of course, to provide hermiticity appropriate conditions should be imposed on 
the coefficients \Math{\alpha_k}. 
\end{enumerate}
Note that other elements of the quantum theory are obtained in the finite background 
in the standard
way. For example, as is well known, the 
\emph{Heisenberg uncertainty principle} follows from the 
\emph{Cauchy-Bunyakovsky-Schwarz inequality}
\begin{equation}
	\inner{A\psi}{A\psi}\inner{B\psi}{B\psi}\geq\cabs{\inner{A\psi}{B\psi}}^2.
	\label{cauchy}
\end{equation}
In fact, in our consideration we deal with the \emph{Cauchy} inequality 
 --- two others extended \eqref{cauchy} to the continuous case.
Obviously, the Cauchy inequality is equivalent to the standard 
property of any probability \Math{\ProbBorn{A\psi}{B\psi}\leq1}.
\section{Permutations, Representations and Numbers}
All transitive actions of a finite group \Math{\wG=\set{\wg_1,\!\ldots\!,\wg_{\wGN}}}
on finite sets can easily be described \cite{HallEn}.
Any such set \Math{\wS=\set{\ws_1,\!\ldots\!,\ws_\wSN}} is 
in one-to-one correspondence with  a set of \emph{right}
\Math{H\backslash\wG} (or \emph{left} \Math{\wG/H})
\emph{cosets} of some subgroup \Math{H\leq\wG}.
The set \Math{\wS} is called a \emph{homogeneous space} of the group \Math{\wG}
(\Math{\wG}-\emph{space} for short).
Action of \Math{\wG} on \Math{\wS} is \emph{faithful}, if the subgroup \Math{H}
does not contain normal subgroups of \Math{\wG}. We can write the action in the form of
permutations
\begin{equation}
	\pi(g)=\dbinom{\ws_i}{\ws_ig}\sim\dbinom{Ha}{Hag},
	\hspace*{20pt}g,a\in{}\wG,~~~i=1,\ldots,\wSN,
	\label{perm}
\end{equation}
or, equivalently, in the form of matrix with entries 0 and 1
\begin{equation}
\pi(g)\rightarrow\regrep(g)=
\Mone{\regrep(g)_{ij}},\text{~~ where~~} \regrep(g)_{ij}=\delta_{\ws_ig,\ws_j};
~~ i,j=1,\ldots,\wSN.
\label{permrepEn}
\end{equation}
Here \Math{\delta_{\alpha,\beta}} is the Kronecker delta on
\Math{\wS}.
Mapping \eqref{permrepEn} is called the \emph{permutation representation}.
\par
Maximal transitive set \Math{\wS} is the set of all elements of the group \Math{\wG}
itself, i.e., the set of cosets of the trivial subgroup \Math{H=\set{\id}}.
The corresponding action and matrix representation are called \emph{regular}. 
One of the central theorems in the representation theory states that 
\emph{any irreducible representation of a finite group is 
contained in the regular representation.}
\par
Representation \eqref{permrepEn} makes sense over any number system with 0 and 1.
A very natural number system is the semi-ring of natural numbers 
\Math{\N=\set{0,1,2,\ldots}.}
With this semi-ring we can attach \emph{counters} to elements of the set 
\Math{\wS}.  These counters (natural numbers) can be interpreted as 
``\emph{multiplicities of occurrences}''
or ``\emph{population numbers}'' of elements \Math{\ws_i} in the state of a system
involving elements from \Math{\wS}. Such state can be represented
by the vector with natural components 
\begin{equation}
	\barket{n} = \Vthree{n_1}{\vdots}{n_{\wSN}}.
\label{natvect}	
\end{equation}
Thus, we come to the representation of the group \Math{\wG} in an 
\Math{\wSN}-dimensional  \emph{module} \Math{\natmod_\wSN} over the semi-ring  \Math{\N}.
Representation \eqref{permrepEn} when applied to vector \eqref{natvect} 
simply permutes its components.
For further development we can turn the module \Math{\natmod_\wSN} into an 
\Math{\wSN}-dimensional  \emph{Hilbert space} \Math{\Hspace_\wSN} by extending 
\Math{\N} to some field.
\par
The main field in the theory
of representations (and hence in the quantum mechanics) is the field of complex numbers \Math{\C}. 
The reason for this choice is simple: the field \Math{\C} is algebraically closed, 
so no complications can be expected in solving characteristic equations and, hence,
in the whole linear algebra. However, the field \Math{\C} is excessively large ---
most of its elements are non-constructive. So let us consider the problem more carefully.
\par
First of all, we do not need to solve \emph{arbitrary} characteristic equations: any
representation is subrepresentation of some permutation representation, and eigenvalues
of any permutation representation are \emph{roots of unity}. This is clear from the
easily calculated \emph{characteristic polynomial}  of permutation matrix \eqref{permrepEn}
\begin{equation*}
\chi_{\regrep(g)}\vect{\lambda}=\det\vect{\regrep(g)-\lambda\idmat}
=\vect{\lambda-1}^{k_1}\vect{\lambda^2-1}^{k_2}\cdots\vect{\lambda^n-1}^{k_n},
\end{equation*}
where \Math{k_i} is the number of cycles of the length \Math{i} in permutation \eqref{perm}.
To provide unitarity of representations we use \emph{square roots} of 
their dimensions as normalizing coefficients. In fact, all \emph{irrationalities} 
(square roots of natural numbers 
and roots of unity) are elements of the same nature ---
they are \emph{cyclotomic integers}, i.e., combinations of roots of unity with
integer coefficients (that can be made natural by using appropriate identities for roots of unity).
\par
Thus, the basic elements of the number system, we are going to construct,
are \emph{natural numbers} and linear combinations (with \emph{natural} coefficients)
 of \emph{roots of unity} 
of some degree 
\Math{\period} depending
on the structure of the group \Math{\wG}. This degree is called \emph{conductor}.
Starting with these basic elements, 
via standard mathematical derivation, we come to a 
\emph{minimal abelian number field} \Math{\NF}
containing these basics. In particular, such field can be computed with the
help of  the computer algebra system \GAP{}\,\cite{gapEn}.
The command
\texttt{\textbf{Field(\textit{gens}\!)}} of  this system 
returns the \emph{smallest} field that contains all elements from
the list of irrationalities \textit{gens}. The field \Math{\NF} can be embedded
into the field \Math{\C}, but we do not need this possibility. Purely algebraic 
properties of \Math{\NF} are sufficient for all manipulations in 
the Hilbert space \Math{\Hspace_\wSN} and its subspaces.
\par
All irrationalities are intermediate elements of quantum description that disappear in the
final expressions for quantum observables --- this is a refinement of the usual 
relationship between the complex and real numbers in the standard quantum mechanics: 
the intermediate values may be complex whereas the final observables are to be real.
\section{Embedding Quantum System into Permutations}
It follows from the above
that any \Math{\adimH}-dimensional representation \Math{\repq}
can be extended to an \Math{\wSN}-dimensional representation
\Math{\widetilde{\repq}} in a Hilbert space \Math{\Hspace_{\wSN}},
in such a way that the representation \Math{\widetilde{\repq}}
corresponds to the \emph{permutation action} of the group \Math{\wG} on
some \Math{\wSN}-element set of entities \Math{\wS=\set{\ws_1,\ldots,\ws_{\wSN}}}.
This means that \Math{\transmatr^{-1}\regrep\transmatr=\widetilde{\repq}},
where \Math{\regrep} is permutation representation \eqref{permrepEn} and \Math{\transmatr} 
is a transformation matrix.
It is clear that \Math{\wSN\geq\adimH}.
\par
The case when \Math{\wSN} is strictly greater than \Math{\adimH}
is most interesting.
In this case the representation has the following
structure
\begin{equation*}
	\transmatr^{-1}\regrep\transmatr
	=\Mthree{\IrrRep{1}}{}{}{}{\repq}{}{}{}{\mathrm{V}}
	\equiv\IrrRep{1}\oplus\repq\oplus\mathrm{V}.
\end{equation*}
Here \Math{\IrrRep{1}} is the trivial one-dimensional representation --- 
an obligatory component of \emph{any} permutation representation.
The component \Math{\mathrm{V}} may be empty.
\par
Clearly, the additional ``hidden parameters''
 --- appearing due to increase 
 of the number of dimension of space in the case  \Math{\wSN>\adimH}
 --- in no way can effect on the data relating to the 
space \Math{\Hspace_{\adimH}} since both \Math{\Hspace_{\adimH}} and
its complement in \Math{\Hspace_{\wSN}}
are invariant subspaces of the extended space \Math{\Hspace_{\wSN}}.
Thus, \emph{any quantum problem} in \Math{\adimH}-dimensional
Hilbert space can be reformulated in terms of permutations of
\Math{\wSN} things.
\par 
With the trivial assumption that the components of state vectors
are \emph{arbitrary} elements of the underlying field  \Math{\NF},
we can set \emph{arbitrary} (e.g., zero) data in the subspace 
\Math{\Hspace_{\wSN-\adimH}} complementary to \Math{\Hspace_{\adimH}}.
In this case we come ---  up to the physically inessential difference
between ``\emph{finite}'' and ``\emph{infinite}'' --- to the standard quantum mechanics 
reformulated in the terms of permutations. 
\par
We can drop this assumption and give more natural meaning to quantum amplitudes.
Let us represent the (quantum) states of the system and apparatus in the permutation
representation by the natural vectors
\Mathh{
\barket{n} = \Vthree{n_1}{\vdots}{n_{\wSN}} \text{~and~}
\barket{m} = \Vthree{m_1}{\vdots}{m_{\wSN}},}
respectively.
In accordance with the Born rule
the probability to fix the system state \Math{\barket{n}} by apparatus tuned to \Math{\barket{m}}
is
\begin{equation}
    \ProbBorn{m}{n}=\frac{\vect{\sum_i{m_i}n_i}^2}{\sum_i{m_i}^2\sum_i{n_i}^2}.
\label{probpEn}
\end{equation}
It is clear that for non-vanishing natural vectors
 \Math{\barket{n}} and \Math{\barket{m}}  expression
\eqref{probpEn} is a rational number strictly greater than zero.
This means, in particular, that it is impossible to observe destructive quantum
interference here.
However, the \emph{destructive interference} of the vectors with natural components
can be observed in the \emph{proper invariant subspaces} of the permutation representation.
We will demonstrate this by a simple example.

\section{Illustrative Example: Group \Math{\SymG{3}} }
\Math{\SymG{3}} is the smallest non-commutative group.
Nevertheless, \Math{\SymG{3}} has important applications in physics.
In particular, it describes the so-called \emph{tribimaximal mixing} in the neutrino oscillations
\cite{HPS02,HS03}. 
The group consists of six elements
having the following representation by permutations
\begin{equation*}
\wg_1=\vect{}\!,~\wg_2=\vect{2,3}\!,~\wg_3=\vect{1,3}\!,~\wg_4=\vect{1,2}\!,
    ~\wg_5=\vect{1,2,3}\!,~\wg_6=\vect{1,3,2}.
\end{equation*}
The group can be generated by many pairs of its elements.
Let us choose, for instance, \Math{\wg_2} and \Math{\wg_6} as generators.
\Math{\SymG{3}} decomposes into the three conjugacy classes
\begin{equation*}
    \class{1}=\set{\wg_1},~~\class{2}=\set{\wg_2,~\wg_3,~\wg_4},
    ~~\class{3}=\set{\wg_5,~\wg_6}.
\end{equation*}
The group \Math{\SymG{3}} has the following character table 
(a compact form to register all irreducible representations)
\begin{equation*}
    \text{\begin{tabular}{c|crr}
    &\Math{\class{1}}&\Math{\class{2}}&\Math{\class{3}}\\\hline
    \Math{\chi_1}&1&1&1\\
    \Math{\chi_2}&1&-1&1\\
    \Math{\chi_3}&2&0&-1
    \end{tabular}\enspace.}
\end{equation*}
In accordance with the physical tradition, we will denote
irreducible representations by their dimensions in bold.
Thus, we have here three irreducible representations \Math{\IrrRep{1}, \IrrRep{1'}} and 
\Math{\IrrRep{2}}  (the last is the only faithful). 
For permutation representations playing an important role in this paper
we will use analogous notation with additional overline.
\par
Matrices of permutation representation of generators are
\begin{equation*}
    P_2=\Mthree{1}{~\cdot}{~\cdot}{\cdot}{~\cdot}{~1}{\cdot}{~1}{~\cdot}
    \text{~and}
    ~P_6=\Mthree{\cdot}{~\cdot}{~1}{1}{~\cdot}{~\cdot}{\cdot}{~1}{~\cdot}.
\end{equation*}
The eigenvalues of \Math{P_2} and \Math{P_6} are \Math{\vect{1, 1, -1}} and
\Math{\vect{1, \runi{}, \runi{}^2}}, respectively; \Math{\runi{}}
is a primitive 3$^d$ root of unity satisfying to the cyclotomic polynomial 
\Math{\Phi_3\vect{\runi{}} = 1+\runi{}+\runi{}^2}.
\par
Since (as is was mentioned above) any permutation representation contains 
one-dimensional invariant subspace with
the basis vector \Math{\vect{1,\ldots,1}^\mathrm{T}}, the only possible structure
of decomposition of permutation representation into irreducible parts is
\Math{\PermRep{3}\cong\IrrRep{1}\oplus\IrrRep{2}} or in the explicit matrix form
\begin{equation}
    \widetilde{U}_j=\Mtwo{\IrrRep{1}}{0}{0}{U_j},~~j =1,\ldots,6,
    \label{S3permqEn}
\end{equation}
where the matrices \Math{U_j} are elements of
the faithful representation \Math{\IrrRep{2}}.
\par
To construct decomposition \eqref{S3permqEn} we should determine matrices \Math{U_j}
and \Math{\transmatr} such that \Math{\widetilde{U}_j=\transmatr^{-1}P_j\transmatr}.
Additionally we impose unitarity on all the matrices.
Clearly, it suffices to perform the procedure only for matrices of generators.
There are different ways to construct decomposition \eqref{S3permqEn}.
If we start with the diagonalization of \Math{P_6}, we come to the following
\begin{align*}
        U_1=\Mtwo{1}{0}{0}{1},~U_2=\Mtwo{0}{\runi{}^2}{\runi{}}{0},
        ~U_3=\Mtwo{0}{\runi{}}{\runi{}^2}{0},\nonumber\\
        ~U_4=\Mtwo{0}{1}{1}{0},
        ~U_5=\Mtwo{\runi{}^2}{0}{0}{\runi{}},~U_6=\Mtwo{\runi{}}{0}{0}{\runi{}^2}.\nonumber
\end{align*}
The transformation matrix (up to inessential degrees of freedom for its
entries) takes the following form
\begin{equation}
\transmatr=\frac{1}{\sqrt{3}}
    \Mthree{1}{1}{\runisymb^2}
     {1}{\runisymb^2}{1}
     {1}{\runisymb}{\runisymb},~~~~
\transmatr^{-1}=\frac{1}{\sqrt{3}}
    \Mthree{1}{1}{1}
     {1}{\runisymb}{\runisymb^2}
     {\runisymb}{1}{\runisymb^2}.
\label{transS3monomial}
\end{equation}
\par
\par
The information about ``quantum behavior'' is encoded, in fact,
in transformation matrices like \eqref{transS3monomial}.
\par
Let \Math{\barket{n} = \Vthree{n_1}{n_2}{n_3}} and
\Math{\barket{m} = \Vthree{m_1}{m_2}{m_3}} be system and apparatus
state vectors in the ``permutation'' basis. Transformation of
these vectors from the permutation to ``quantum'' basis with the
help of \eqref{transS3monomial} leads to
\begin{align*}
    \barket{\widetilde{\psi}}=\transmatr^{-1}\barket{n}
    &=\frac{1}{\sqrt{3}}\Vthree{n_1+n_2+n_3}
    {n_1+n_2\runisymb+n_3\runisymb^2}{n_1\runisymb+n_2+n_3\runisymb^2},\\
    \barket{\widetilde{\phi}}=\transmatr^{-1}\barket{m}
    &=\frac{1}{\sqrt{3}}\Vthree{m_1+m_2+m_3}
    {m_1+m_2\runisymb+m_3\runisymb^2}{m_1\runisymb+m_2+m_3\runisymb^2}.
\end{align*}
Projections of the vectors onto two-dimensional invariant subspace
 are:
\begin{equation*}
    \barket{\psi} = \frac{1}{\sqrt{3}}\Vtwo{n_1+n_2\runisymb+n_3\runisymb^2}
    {n_1\runisymb+n_2+n_3\runisymb^2},~~~~
    \barket{\phi} = \frac{1}{\sqrt{3}}\Vtwo{m_1+m_2\runisymb+m_3\runisymb^2}
    {m_1\runisymb+m_2+m_3\runisymb^2}.
\end{equation*}
Constituents of Born's probability \eqref{BornEn} for the two-dimensional subsystem
 are
\begin{equation}
    \inner{\psi\!}{\!\psi}=\invarQ{3}{n}{n}-\frac{1}{3}\invarL{3}{n}^2,
\label{Born2den1En}
\end{equation}
\begin{equation}
    \inner{\phi\!}{\!\phi}=\invarQ{3}{m}{m}-\frac{1}{3}\invarL{3}{m}^2,
\label{Born2den2En}
\end{equation}
\begin{equation}
\cabs{\inner{\phi\!}{\!\psi}}^2=
\vect{\invarQ{3}{m}{n}-\frac{1}{3}\invarL{3}{m}\invarL{3}{n}}^2,
\label{Born2numEn}
\end{equation}
where \Math{\invarL{\wSN}{n}=\sum\limits_{i=1}^{\wSN}n_i} and
\Math{\invarQ{\wSN}{m}{n}=\sum\limits_{i=1}^{\wSN}m_in_i} are linear and
quadratic permutation invariants, respectively.
\par
Note that:
\begin{enumerate}
    \item Expressions \eqref{Born2den1En}--\eqref{Born2numEn} are combinations of the
    \emph{invariants of permutation representation}.
    \item
Expressions \eqref{Born2den1En} and \eqref{Born2den2En} are always
positive rational numbers for \Math{\barket{n}} and
\Math{\barket{m}} with different components.
    \item
Conditions for \emph{destructive quantum interference} ---
vanishing Born's probability ---    are determined by the equation
\Mathh{3\vect{m_1n_1+m_2n_2+m_3n_3}-\vect{m_1+m_2+m_3}\vect{n_1+n_2+n_3}=0.}
This equation has infinitely many solutions in natural numbers.
An example of such a solution is:
\Math{{\barket{n} = \Vthree{1}{1}{2},~~\barket{m} = \Vthree{1}{3}{2}}}.
\end{enumerate}
Thus, we have obtained
essential features of quantum behavior from ``permutation dynamics''
and ``natural'' interpretation \eqref{natvect} of quantum amplitude
by a simple transition to invariant subspaces.
\par
Any permutation representation contains 
\Math{\vect{\wSN-1}}-dimensional invariant subspace. The inner
product in this subspace can be expressed in terms of the
permutation invariants by the formula
\Mathh{\inner{\phi\!}{\!\psi}=
\invarQ{\wSN}{m}{n}-\frac{1}{\wSN}\invarL{\wSN}{m}\invarL{\wSN}{n}.}
The identity
\Math{\displaystyle\invarQ{\wSN}{n}{n}-\frac{1}{\wSN}\invarL{\wSN}{n}^2\equiv
\frac{1}{\wSN^2}\sum\limits_{i=1}^\wSN\vect{\invarL{\wSN}{n}-\wSN{}n_i}^2}
shows explicitly that \Math{\inner{\psi\!}{\!\psi}>0} for
\Math{\barket{n}} with different components \Math{n_i}. This inner
product does not contain irrationalities for natural
\Math{\barket{n}} and \Math{\barket{m}}. 
\section{Icosahedral Group \Math{\AltG{5}}}
\begin{figure}
\centering
\includegraphics[width=0.65\textwidth]{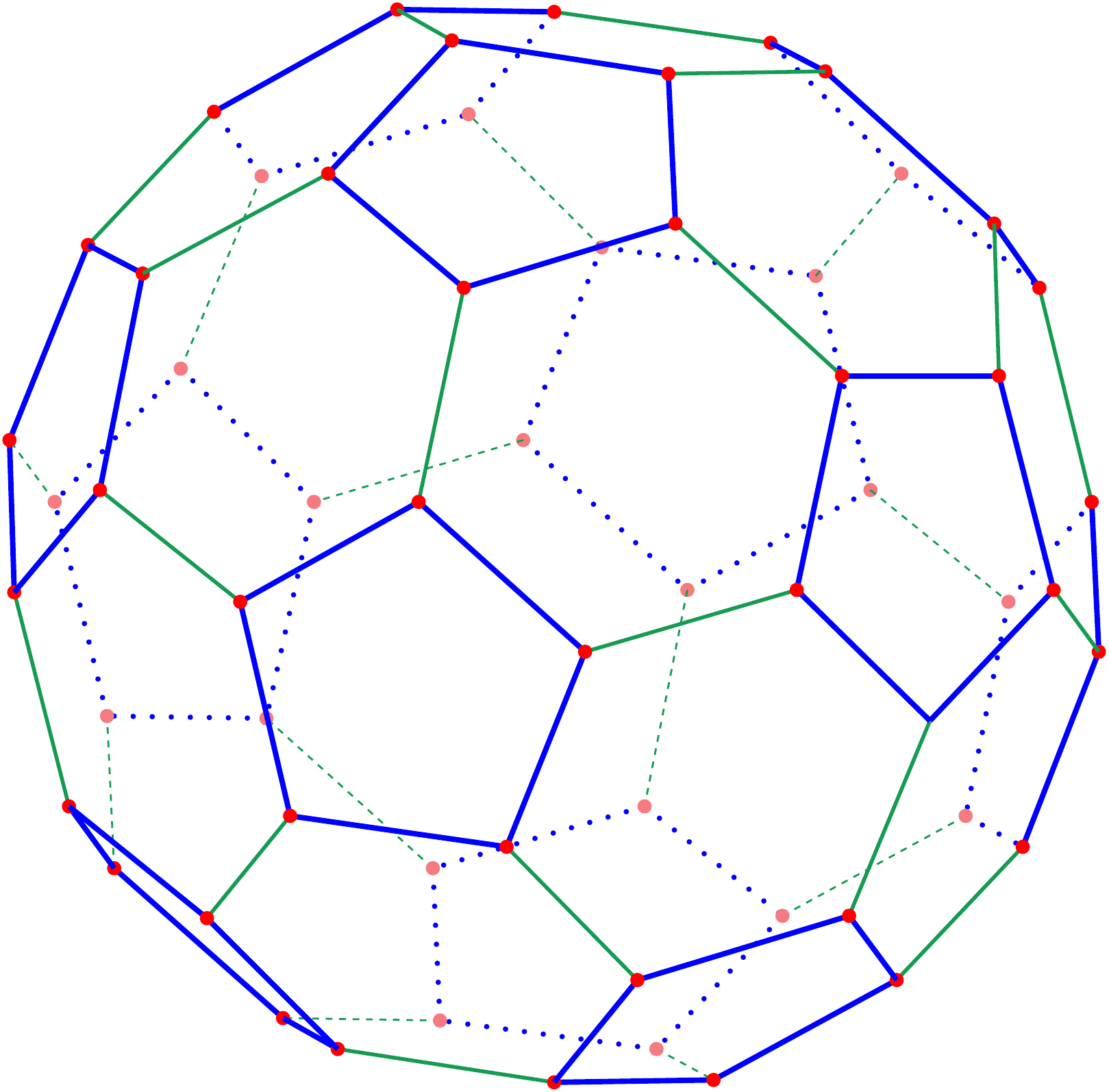}
\caption{The Cayley graph of \Math{\AltG{5}}. 
The pentagons, hexagons and the links connecting adjacent pentagons correspond to 
the relators \Math{a^5}, 
\Math{\vect{ab}^3} and  \Math{b^2} in presentation 
\eqref{present}, respectively.}
\label{bucky}
\end{figure}

The icosahedral group \Math{\AltG{5}} --- the smallest (it consists of 60 elements) 
simple non-commutative group
--- plays an important role in mathematics and applications.
F. Klein devoted a whole book to it \cite{Klein}.
  In the physical literature the group is often denoted 
as \Math{\Sigma\!\vect{60}}.
This group
has a ``physical incarnation'': the fullerene \Math{C_{60}} carbon molecule
(``{buckyball}'') has the structure of the Cayley graph of  \Math{\AltG{5}} 
(see Fig. \ref{bucky}).
This is clear from the following \emph{presentation} of \Math{\AltG{5}} by 
\emph{generators} and \emph{relators} (products of generators that are 
equal to the group identity)
\begin{equation}
	\AltG{5}\cong\left\langle{}a, b\mid{}a^5,
	b^2, \vect{ab}^3\right\rangle.
	\label{present}
\end{equation}
\par
Let us consider the action of \Math{\AltG{5}} on the set \Math{\wS_{12}} 
of icosahedron vertices. 
\begin{figure}
\centering
\includegraphics[width=0.65\textwidth]{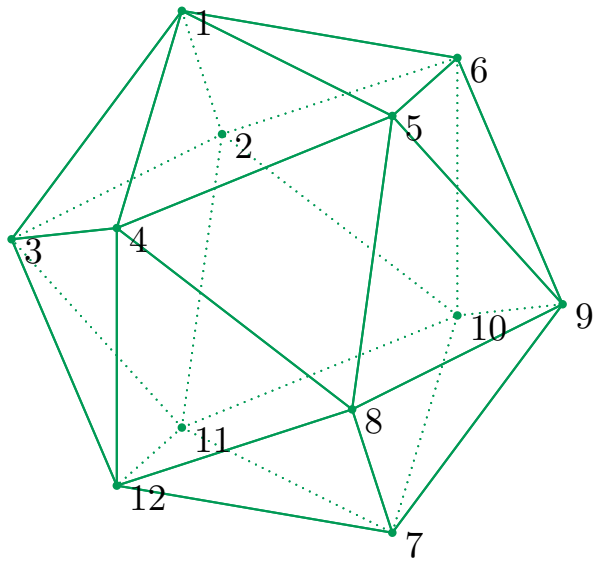}
\caption{The icosahedron. Pairs of opposite vertices form the system of imprimitivity.}
\label{ico}
\end{figure}
This action is transitive but \emph{imprimitive}, i.e., 
there is non-trivial partition of  \Math{\wS_{12}} invariant under the action. 
This partition --- called
\emph{system of imprimitivity} (or \emph{block system}) --- is the
following 
\Mathh{\set{\mid{}B_1\mid\cdots\mid{}B_i\mid\cdots\mid{}B_6\mid}\equiv
\set{\mid1,7\mid\cdots\mid{}i,i+6\mid\cdots\mid6,12\mid}} 
where, the vertex numbering of Fig. \ref{ico} is assumed. Each block \Math{B_i}
is a pair of two opposite vertices of the icosahedron and \Math{\AltG{5}} permutes
the blocks amongst themselves as whole entities. We shall denote the 
correspondence (``complementarity'') between opposite vertices by the symbol \Math{\Oppbare}, 
i.e., if 
\Math{B_i=\set{p,q}} then \Math{q=\Opp{p}} and \Math{p=\Opp{q}}.
For the vertex numbering depicted in Fig. \ref{ico} the complementarity can be 
expressed by the formula \Math{\Opp{p}\equiv1+\vect{p+5\mod12}}. 
\par
The permutation representation of the action of \Math{\AltG{5}} on the vertices of 
icosahedron has the following decomposition into irreducible components
\begin{equation}
	\PermRep{12}\cong\IrrRep{1}\oplus\IrrRep{3}\oplus\IrrRep{3'}\oplus\IrrRep{5}
	\text{~~or~~}\transmatr^{-1}\vect{\PermRep{12}}\transmatr
	=\IrrRep{1}\oplus\IrrRep{3}\oplus\IrrRep{3'}\oplus\IrrRep{5}.
	\label{decoico}
\end{equation}
Note, that the group \Math{\AltG{5}} has three \emph{primitive} actions on sets 
with 5, 6 and 10 elements. The corresponding permutation representations have the
following decompositions
\Mathh{\PermRep{5}\cong\IrrRep{1}\oplus\IrrRep{4},
~\PermRep{6}\cong\IrrRep{1}\oplus\IrrRep{5},
~\PermRep{10}\cong\IrrRep{1}\oplus\IrrRep{4}\oplus\IrrRep{5}.}
With the notations\\[3pt] 
\Math{\displaystyle\phi=\frac{1+\sqrt{5}}{2}} --- the \emph{``golden ratio''},~~ 
\Math{\displaystyle\alpha=\frac{\phi}{4}\sqrt{10-2\sqrt{5}}},\\
\Math{\displaystyle\beta=\frac{\sqrt{5}\sqrt{10-2\sqrt{5}}}{20}},
\Math{\displaystyle\gamma=\frac{\sqrt{3}}{8}\vect{1-\frac{\sqrt{5}}{3}}},~~
\Math{\displaystyle\delta=-\frac{\sqrt{3}}{8}\vect{1+\frac{\sqrt{5}}{3}}}\\[3pt]
a particular form of unitary transformation 
matrix \Math{\transmatr} from \eqref{decoico} can be written as
\Mathh{
\transmatr=
\bmat
\frac{\sqrt{3}}{6}&\alpha&\beta&0&\alpha&\beta&0&\frac{1}{4}&-\frac{1}{2}&0&0&\frac{\sqrt{15}}{12}
\\
\frac{\sqrt{3}}{6}&0&\alpha&\beta&-\beta&0&\alpha&-\frac{\phi}{4}&0&-\frac{1}{2}&0&\gamma
\\
\frac{\sqrt{3}}{6}&\beta&0&\alpha&0&-\alpha&-\beta&\frac{\phi-1}{4}&0&0&-\frac{1}{2}&\delta
\\
\frac{\sqrt{3}}{6}&0&\alpha&-\beta&-\beta&0&-\alpha&-\frac{\phi}{4}&0&\frac{1}{2}&0&\gamma
\\
\frac{\sqrt{3}}{6}&-\beta&0&\alpha&0&\alpha&-\beta&\frac{\phi-1}{4}&0&0&\frac{1}{2}&\delta
\\
\frac{\sqrt{3}}{6}&\alpha&-\beta&0&-\alpha&\beta&0&\frac{1}{4}&\frac{1}{2}&0&0&\frac{\sqrt{15}}{12}
\\
\frac{\sqrt{3}}{6}&0&-\alpha&\beta&\beta&0&\alpha&-\frac{\phi}{4}&0&\frac{1}{2}&0&\gamma
\\
\frac{\sqrt{3}}{6}&\beta&0&-\alpha&0&-\alpha&\beta&\frac{\phi-1}{4}&0&0&\frac{1}{2}&\delta
\\
\frac{\sqrt{3}}{6}&-\alpha&\beta&0&\alpha&-\beta&0&\frac{1}{4}&\frac{1}{2}&0&0&\frac{\sqrt{15}}{12}
\\
\frac{\sqrt{3}}{6}&-\alpha&-\beta&0&-\alpha&-\beta&0&\frac{1}{4}&-\frac{1}{2}&0&0&\frac{\sqrt{15}}{12}
\\
\frac{\sqrt{3}}{6}&0&-\alpha&-\beta&\beta&0&-\alpha&-\frac{\phi}{4}&0&-\frac{1}{2}&0&\gamma 
\\
\frac{\sqrt{3}}{6}&-\beta&0&-\alpha&0&\alpha&\beta&\frac{\phi-1}{4}&0&0&-\frac{1}{2}&\delta
\emat.
}
Note that the standard computer algebra systems like \emph{Maple} or
\emph{Mathematica} can not handle such matrices since they can not simplify 
complicated expressions with irrationalities (especially with nested roots) properly.
But if one rewrites the matrix entries as elements of suitable abelian number field
\Math{\NF}, then 
the problem of simplification is reduced to a simple one-variable polynomial algebra 
\textit{modulo} corresponding cyclotomic polynomial.  
\par
The inner products in the invariant subspaces can be expressed in terms of
permutation invariants as follows:
\begin{align}
\displaystyle\inner{\Phi_{\IrrRep{1}}}{\Psi_{\IrrRep{1}}}&=
\frac{1}{12}\invarL{12}{m}\invarL{12}{n},
\\
\displaystyle\inner{\Phi_{\IrrRep{3}}}{\Psi_{\IrrRep{3}}}&=
\frac{1}{20}\vect{5\invarQ{12}{m}{n}-5\invar{A}{m}{n}
+\sqrt{5}\vect{\invar{B}{m}{n}-\invar{C}{m}{n}}},\label{inn3}
\\
\displaystyle\inner{\Phi_{\IrrRep{3'}}}{\Psi_{\IrrRep{3'}}}&=
\frac{1}{20}\vect{5\invarQ{12}{m}{n}-5\invar{A}{m}{n}
-\sqrt{5}\vect{\invar{B}{m}{n}-\invar{C}{m}{n}}},\label{inn3'}
\\
\displaystyle\inner{\Phi_{\IrrRep{5}}}{\Psi_{\IrrRep{5}}}&=
\frac{1}{12}\vect{5\invarQ{12}{m}{n}+5\invar{A}{m}{n}
-\invar{B}{m}{n}-\invar{C}{m}{n}},
\end{align}
where
\begin{align}
\displaystyle\invar{A}{m}{n}&
=\invar{A}{n}{m}=\sum\limits_{k=1}^{12}m_kn_{\Opp{k}},\label{Amn}\\
\displaystyle\invar{B}{m}{n}&
=\invar{B}{n}{m}=\sum\limits_{k=1}^{12}m_k\!\!\sum\limits_{i\in\Nei{k}}\!\!n_{i},\label{Bmn}\\
\displaystyle\invar{C}{m}{n}&
=\invar{C}{n}{m}=\sum\limits_{k=1}^{12}m_k\!\!\!\sum\limits_{i\in\Nei{\Opp{k}}}\!\!\!\!\!n_{i}.
\label{Cmn}
\end{align}
In formulas \eqref{Bmn} and \eqref{Cmn}  \Math{\Nei{k}} denotes the ``neighborhood'' 
of the icosahedron vertex \Math{k}, 
i. e., the set of vertices adjacent to \Math{k}. 
For example,  \Math{\Nei{1}=\set{2,3,4,5,6}} in Fig. \ref{ico}. 
Quadratic  invariants \eqref{Amn}--\eqref{Cmn} are not independent. There is an  
identity among them:
\Mathh{\displaystyle\invar{A}{m}{n}+\invar{B}{m}{n}+\invar{C}{m}{n}+\invarQ{12}{m}{n}
=\invarL{12}{m}\invarL{12}{n}.}
Inner products \eqref{inn3} and  \eqref{inn3'} lead to the mentioned above conceptual 
difficulty with probability 
when considered separately. Born's probabilities computed separately for the 
representations \Math{\IrrRep{3}} and \Math{\IrrRep{3'}} contain irrationalities.
This contradicts the frequency interpretation of probability for finite sets.
Of course, this is a consequence of the imprimitivity: one can not move a vertex of the
icosahedron without simultaneous moving of its complement in the block. 
To resolve the contradiction we should consider the complementary  
representations \Math{\IrrRep{3}} and \Math{\IrrRep{3'}} together.
The inner product for the representation \Math{\IrrRep{3}\oplus\IrrRep{3'}}
takes the form 
\begin{equation*}
\displaystyle\inner{\Phi_{\IrrRep{3\oplus3'}}}{\Psi_{\IrrRep{3\oplus3'}}}=
\frac{1}{2}\vect{\invarQ{12}{m}{n}-\invar{A}{m}{n}}.	
\end{equation*}
This inner product always gives rational Born's probabilities for vectors with natural
``population numbers''.   
\section*{Conclusions}
Let us summarize the main ideas of the paper
\begin{enumerate}
\item 
Quantum mechanics is, in fact, an \alert{a priori mathematical scheme} based
on the fundamental impossibility to trace identity of indistinguishable objects
in their evolution --- some kind of \emph{``calculus of indistinguishables''}.
\item
\alert{Any} quantum mechanical problem can be \alert{reduced to permutations}.
\item
\emph{Quantum interferences} are appearances observable in the invariant subspaces 
of permutation representations --- they can be expressed in terms of 
the \emph{permutation invariants}.
\item
Natural interpretation of \alert{quantum amplitudes} (\emph{``waves''}) as vectors of 
\emph{``population numbers''} of underlying 
entities (\emph{``particles''}) leads to the \alert{rational} quantum probabilities    
--- in accordance with the \alert{frequency interpretation} of probability 
for finite sets.
\end{enumerate}
The idea of natural quantum amplitudes requires verification.
If it is valid, then quantum phenomena in different invariant subspaces
reveal different manifestations --- visible in different \emph{``observational settings''}
--- of a single process of permutations of underlying things.
One has to interpret the data corresponding to different invariant subspaces. 
For example, the trivial one-dimensional subrepresentation
contained in any permutation representation can be interpreted as the 
\emph{``counter of particles''}:~ the permutation invariant \Math{\invarL{\wSN}{n}} corresponding to 
this subrepresentation is the total 
number of particles. Interpretation of data in other invariant subspaces requires
 careful  study.
\paragraph{\bf{}Acknowledgment.}
The author is grateful to  Yuri Blinkov and Vladimir Gerdt for fruitful 
discussions and for their help in preparing the paper.
The work was partially supported by the grants 01-01-00200 from the Russian 
Foundation for Basic
Research and 3810.2010.2 from the Ministry of Education and Science of
the Russian Federation.
\par

\end{document}